\documentclass{pasa}%

\usepackage{graphicx}
\usepackage[dvipsnames]{xcolor}
\usepackage{tablefootnote}

\title[High Cadence Optical Transient Searches]{High Cadence Optical Transient Searches using Drift Scan Imaging II: Event Rate Upper Limits on Optical Transients of Duration $<$21 ms and Magnitude $<$6.6}

\author[]{Steven Tingay and Wynand Joubert
\affil{International Centre for Radio Astronomy Research, Curtin University, Bentley, WA 6102, Australia}%
}%

\jid{PASA}
\doi{10.1017/pas.\the\year.xxx}
\jyear{\the\year}

\usepackage{aas_macros}
\usepackage{hyperref} 
\hypersetup{colorlinks,citecolor=blue,linkcolor=blue,urlcolor=blue}

\begin{document}

\begin{frontmatter}
\maketitle

\begin{abstract}
We have realised a simple prototype system to perform searches for short timescale optical transients, utilising the novel drift scan imaging technique described by \citet{2020PASA...37...15T}.  We used two coordinated and aligned cameras, with an overlap field-of-view of approximately 3.7 sq. deg., to capture over $34000 \times 5$ second images during approximately 24 hours of observing.  The system is sensitive to optical transients, due to an effective exposure time per pixel of 21 ms, brighter than a V magnitude of 6.6.  In our 89.7 sq. deg. hr of observations we find no candidate astronomical transients, giving an upper limit to the rate of these transients of 0.8 per square degree per day, competitive with other experiments of this type.  The system is triggered by reflections from satellites and various instrumental effects, which are easily identifiable due to the two camera system.  The next step in the development of this promising technique is to move to a system with larger apertures and wider fields of view.
\end{abstract}

\end{frontmatter}

\section{INTRODUCTION }
\label{sec:introduction}
In order to explore a rarely entered region of optical transient search space, \citet{2020PASA...37...15T} (hereafter Paper I) presented a novel method to obtain high time resolution measurements using long exposure images, mitigating the significant data volume issues encountered when searching for transients with duration $<<$1 sec.  This work was motivated by the discovery of astrophysical sources of transient radio emission on millisecond timescales, for example Fast Radio Bursts (FRBs; \citet{2018NatAs...2..845B}), and the need to obtain multi-wavelength data for such objects.  

The reader is referred to Paper I for a detailed description of the motivation for our work, as well a review of previous experiments conducted at $<$1 sec time resolution and/or searching for FRB counterparts at optical wavelengths.  In addition, recently \citet{2020arXiv200506273A} used the Zwicky Transient Facility (ZTF) to search for optical emission associated with the repeating FRB 180916.J0158$+$65, localised to a spiral galaxy 149 Mpc distant.  They found no detection above 3$\sigma$ using 30 s duration exposures over 5.69 hours of observations, but were able to place a limit on the ratio of optical to radio fluences of $<$200 (90\% confidence).  \citet{2020arXiv200506273A} also provide an excellent review of the current state of FRB detections.

The methods described in Paper I involve the drift of the sky across the field-of-view of a sensor while undertaking a long exposure, achieving high time resolution by virtue of the fact that an element of sky occupies any given sensor pixel for a short period of time; short duration signals are localised in the drift scan image, whereas persistent signals form a trail.  The demonstration of this method, using a simple pre-prototype system composed of commodity components, showed that it could achieve an effective cadence of 21 ms, a field-of-view of 4.4 sq. deg., and a limiting sensitivity in 21 ms of V=6.6 mag.  While high time resolution is achieved efficiently because the long exposures limit the data rate from images, as discussed in Paper I, localisation in astronomical coordinates is compromised in the drift direction (Right Ascension), making this an experiment most suited to determining the existence and rate of short timescale transients.

Detections with the pre-prototype were dominated by the presence of cosmic ray hits on the sensor, being indistinguishable from signals of astrophysical interest when using only a single camera.  A proposed enhancement of the pre-prototype system, described in Paper I, was the addition of a second, identical camera, pointed at the same area of sky, for the purpose of differentiating between cosmic ray hits and astrophysically interesting events.

In the current paper we implement this enhancement and utilise two identical and synchronised cameras in order to undertake a survey of significant duration with the performance parameters described above, eliminating cosmic ray hits as a source of candidate events.  In \S \ref{sec:hardware}, we describe the enhanced system utilised here.  In \S \ref{sec:obsdataproc} we describe the observations performed and the data processing undertaken.  In \S \ref{res}, we describe the results of our survey and place the results in the context of astrophysical expectations and prior similar work.

\section{SYSTEM HARDWARE}
\label{sec:hardware}
The system utilised for this work is as described in Paper I, with the enhancement described in \S5.1 of Paper I, the addition of a second identical camera and lens to provide a mechanism to identify false candidates due to cosmic ray hits on the sensor.  Briefly, the cameras are Canon 600D digital cameras equipped with Samyang 500 mm focal length, f/6.3 (aperture diameter of approximately 79 mm) lenses.  The Canon 600D has a sensor size of 5190$\times$3461 pixels and a pixel size of 4.29$\mu$m, giving an approximate 2.5$^{\circ}\times$1.7$^{\circ}$ field of view.

\begin{figure}[!ht]
\begin{center}
\includegraphics[width=0.45\textwidth]{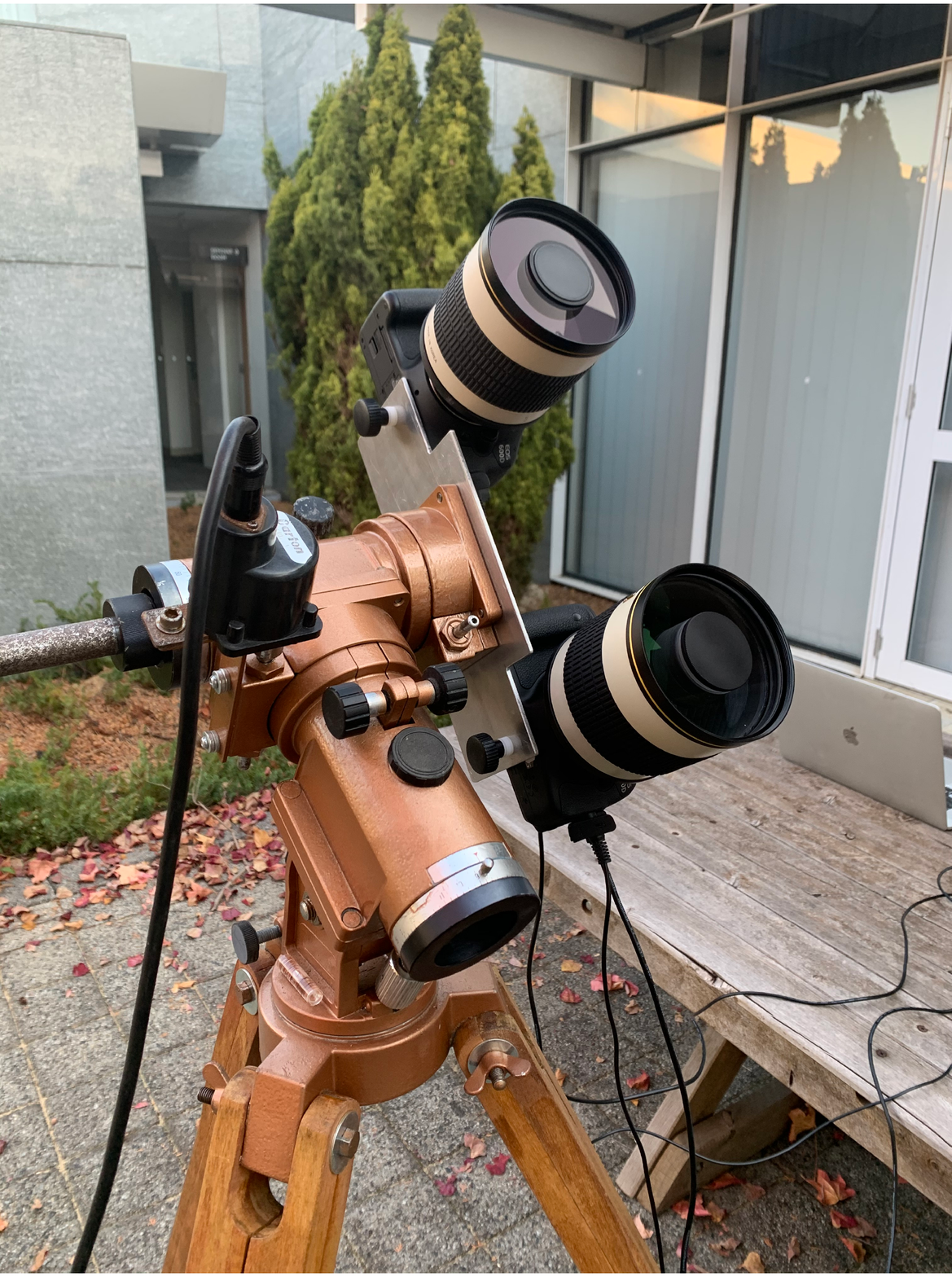}
\caption{The two cameras mounted side-by-side for simultaneous operation.  Camera A is at top, Camera B at bottom.}
\label{mount}
\end{center}
\end{figure}

The two cameras were mounted side-by-side on the same mount as used in Paper I, as shown in Figure \ref{mount}.

In order to configure the two cameras for simultaneous operation, an Esper TriggerBox\footnote{ https://www.esperhq.com/product/multiple-camera-trigger-triggerbox/} was used to schedule and synchronise the timing of the exposures for the two cameras.  The onboard clocks of the two cameras were synchronised in order to obtain the same time stamps on simultaneous exposures.

\section{Observations and Data Processing}
\label{sec:obsdataproc}
Observations took place on the dates and times listed in Table \ref{tab1} from two suburban Perth locations.  The observations of 20200520, 20200616, and 20200722 were undertaken from 115$^{\circ}$:53':20'' E; 31$^{\circ}$:59':40'' S.  The remaining observations were undertaken from 115$^{\circ}$:54':12'' E; 32$^{\circ}$:03':09'' S. The observations of 20200520 constituted a short, successful first commissioning of the system, that yielded science quality data.  The other entries in Table \ref{tab1} represent the ongoing program of observations.  All observations were made on cloud-free nights.  Separate entries for the same night of observation exist in Table \ref{tab1}, corresponding to changes of the memory cards in the cameras, possibly perturbing the camera alignment and therefore requiring separate astrometric calibrations.

\begin{table*}[h!]
\caption{Observation log}
\centering
\begin{tabular}{|c|c|c|c|c|c|c|} \hline \hline
Date    &Time Range$^{a}$&Cam.&Image Centre$^{b}$          &Image Centre$^{b}$      &Rotation$^{b}$         &Field of View$^{c}$ \\
(yyyymmdd)&(UT)/$\Delta$T (hr)            &      &RA (J2000)                  &Dec (J2000)             &$^{\circ}$ EoN         &(sq. deg.)           \\ \hline
20200520&11:56$-$12:21   &A     &187.627                     &$-$4.603                &93.9                   &3.8                  \\  
&0.30                &B     &187.825                     &$-$4.569                &93.5                   &                     \\ \hline
20200616&10:26$-$11:35&A&173.024&$-$6.050&92.0&3.8 \\
&1.0&B&173.208&$-$6.073&92.1& \\ \hline
20200616&11:51$-$13:04&A&198.045&$-$5.992&92.0&3.9 \\
&1.0&B&198.216&$-$6.018&92.1& \\ \hline
20200722&10:20$-$11:45&A&212.556&$-$7.338&102.0&3.7 \\
&1.0&B&212.718&$-$7.509&102.0& \\ \hline 
20200722&11:55$-$12:22&A&202.692&$-$10.317&101.0&3.7 \\
&0.45&B&205.878&$-$10.496&101.0& \\ \hline
20200913&10:47$-$12:06&A&250.391&$-$6.228&91.9&3.8 \\
&1.0&B&250.561&$-$6.131&92.0& \\ \hline
20200913&12:13$-$12:45&A&273.113&$-$6.514&91.9&3.8 \\
&0.40&B&273.277&$-$6.415&92.0& \\ \hline
20200915&12:10$-$13:54&A&282.630&$-$8.508&95.3&3.6 \\
&0.93&B&282.805&$-$8.304&95.5& \\ \hline 
20200915&13:58$-$15:00&A&294.218&$-$4.096&94.9&3.6 \\
&1.0&B&294.398&$-$3.897&95.1& \\ \hline
20200915&15:17$-$16:32&A&311.454&$-$4.359&94.7&3.6 \\
&1.0&B&311.636&$-$4.161&94.9& \\ \hline
20200915&16:41$-$17:30&A&335.045&$-$3.274&94.9&3.6 \\
&0.59&B&335.231&$-$3.082&95.0& \\ \hline
20200918$^{d}$&10:53$-$16:53&A&253.165&$-$7.606&92.2&3.9 \\
&3.83&B&253.354&$-$7.594&92.4& \\ \hline
20200922&10:54$-$15:48&A&254.455&$-$2.348&103.0&3.7 \\
&4.0&B&254.680&$-$2.270&103.0& \\ \hline
20201010&11:50$-$12:45&A&298.704&$-$6.967&89.6&3.6 \\
&0.71&B&298.977&$-$6.973&89.9& \\ \hline
20201011&10:44$-$15:50&A&298.598&$-$6.078&92.9&3.7 \\
&3.83&B&298.857&$-$6.067&93.0& \\ \hline
20201012&11:21$-$15:40&A&298.625&$-$7.685&92.7&3.7 \\
&2.99&B&298.880&$-$7.667&92.9& \\ \hline

\end{tabular} \\
$^{a}$ Applies to Cameras A and B; $^{b}$ For astrometry frames; $^{c}$ Overlapping fields of view of Cameras A and B; $^{d}$ We obtained 128 GB SD cards, allowing longer observing sessions.
\label{tab1}
\end{table*}

As described in Paper I, for each observation, a set of dark frames were acquired to produce a Bad Pixel Map for each camera, used to remove hot pixels from the subsequent science frames.  For each camera, the Bad Pixel Map eliminated less than 1\% of the pixels (typically 0.7\% - 0.9\%).

As described in Paper I, the cameras were pointed immediately south of the celestial equator, near the meridian, in order to avoid the belt of geosynchronous satellites that appear north of the equator from our observing location.  The cameras were focused using a bright star.

A unit of observation is defined whereby the cameras are pointed to the west of the meridian and then actively driven in the anti-sidereal direction at a rate across the sensor of 71''/s (as described in Paper I).  The unit of observation was variable across different nights and even within nights, and generally got longer as we built confidence with the system and data processing.  On some nights, the avoidance of passing isolated clouds in some parts of the sky led to variable units of observation.  The minimum unit used was three minutes and the maximum unit was 30 minutes.

During a unit of observation, the cameras synchronously obtain a sequence of 5 s images, which are stored on board the cameras in the native Canon CR2 format.  At the end of a unit of observation, the pointing direction is reset to the west of the meridian and the unit of observation is reset.  The images are transferred from the cameras to a laptop computer for processing.  For the observations described in Table \ref{tab1}, a total of 17,301 images per camera were obtained over 10 nights of observation, resulting in a total observation time of 24.03 hours and a total data volume of approximately 670 GB.

\subsection{Alignment of cameras}
\label{align}
For each observation, a series of short exposures were obtained to assess the alignment of the two cameras.  The resulting images were uploaded to www.astrometry.net, to solve for the astrometry, as described in Paper I.  The astrometry comparison between the two cameras for each observation is shown in Table \ref{tab1}.  For all observations, the plate scales for Camera A and Camera B are 1.78''/pix and 1.77''/pix, respectively.  The field of view for Cameras A and B are $2.56^{\circ}\times1.71^{\circ}$ and $2.55^{\circ}\times1.70^{\circ}$, respectively.

The astrometric solution from astrometry.net returns a FITS file describing the World Coordinate System (WCS) for each camera, which can then be used to map the pixel locations between the two cameras.

\begin{figure}[!ht]
\begin{center}
\includegraphics[width=0.45\textwidth]{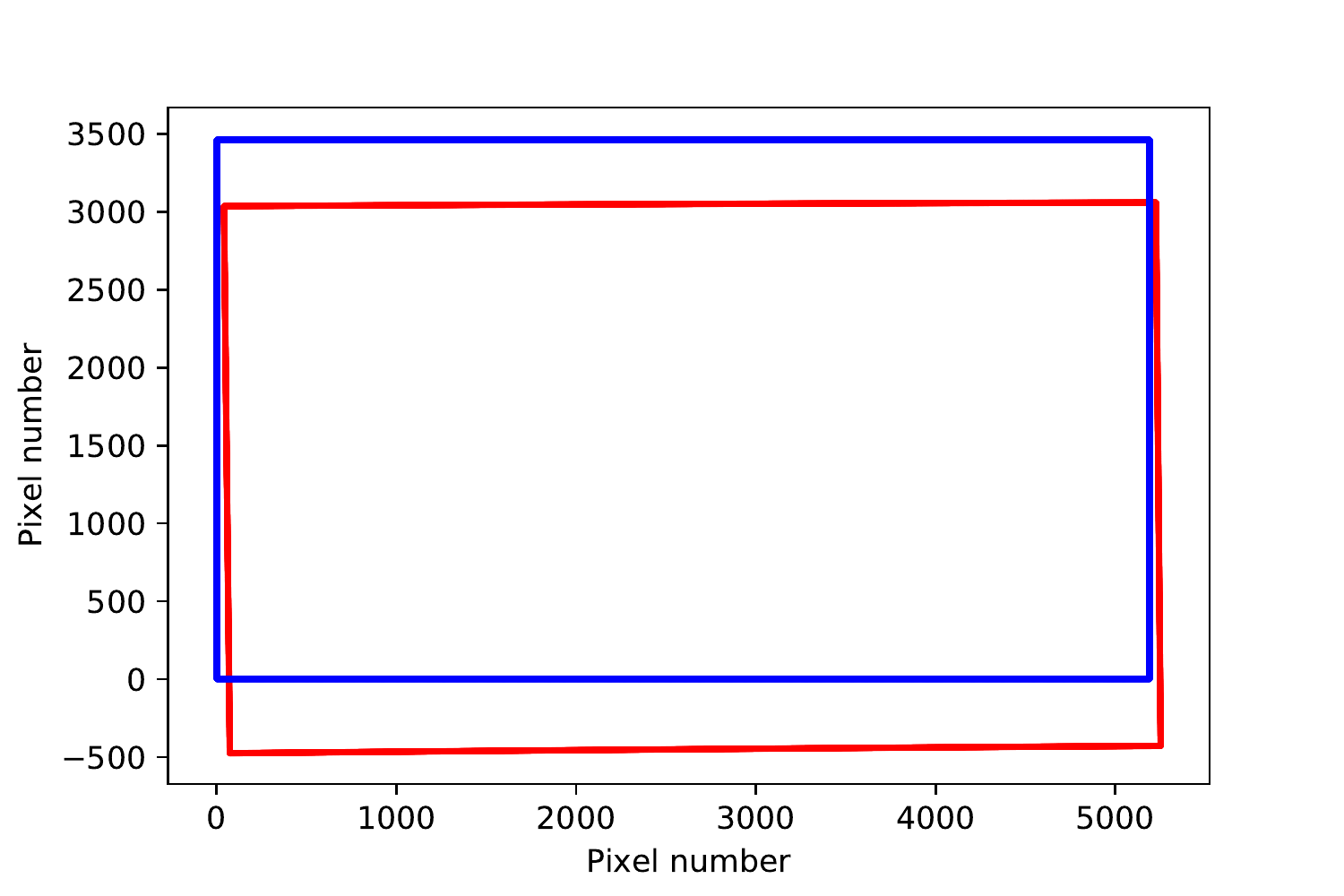}
\caption{Example field of view overlap, for the commissioning observations of 20200520.  Blue frame is for Camera A, red frame is for Camera B.  In this figure, east is up and north is left.}
\label{overlap}
\end{center}
\end{figure}

The overlap in the fields-of-view for the two cameras is calculated using the WCS information and is listed in Table \ref{tab1}.  An example of the field-of-view overlap (which may be different each time the system is set up, in general) is shown in Figure \ref{overlap}, for the commissioning observations of 20200520.

The transfer of the astrometry between the two cameras was verified by calculating the RA and Dec values for the stars detected as part of the astrometric solution for Camera A (using the Camera A WCS) and converting them to $x$ and $y$ values for Camera B (using the Camera B WCS).  We compared these $x$ and $y$ values to the corresponding values for the stars detected as part of the Camera B astrometric solution.  We found agreement at better than 30 pixels for all stars.  Thus, we adopt a matching radius of 30 pixels when searching for candidate transient events in our data.

\subsection{Data Processing}
\label{proc}
As described in Paper I, within the Nebulosity software package\footnote{https://www.stark-labs.com/nebulosity.html} the relevant Bad Pixel Map was applied to the sequence of images for each camera for each observation, images were demosaiced and pixels squared, frames were converted to monochrome, and then saved as FITS files.

In Paper I, an extensive analysis of the drift scan images was undertaken, to verify the image processing, astrometry, and photometry, establishing the drift scan method.  Paper I also undertook a comprehensive analysis of the candidate signals, concluding that they were overwhelmingly likely to be due to cosmic ray hits on the sensor.  This fact motivated the current configuration, with the use of two cameras to identify and eliminate these false candidates; cosmic rays will {\bf rarely} occur at the same celestial coordinates in the two cameras, or at the same times.

Compared to the data processing undertaken in Paper I, the use of two cameras greatly simplifies the data processing required here.  As described in Paper I, difference images were formed for all images from cameras A and B, and searched for candidate signals above a threshold of 10 times the RMS in each difference image, and with the distinctive signature in the difference images of a high positive pixel alongside a high negative pixel in a column, in the correct order (as described in Paper I).  These criteria are designed to detect point-like signals in the images, but will also detect signals that extend over significant numbers of pixels (in any direction) if sufficiently strong, as noted below.  The criteria are sensitive to a wide range of signal types in the images, thus do not narrowly restrict the signal types that are passed for coincidence checking.

Using the mapping between camera pixels derived in \S \ref{align}, we take the candidate list for camera A and map their pixel locations to pixel locations for camera B, using the WCS solutions for each camera.  If a candidate from the camera B list occurs within 30 pixels of the location derived from camera A, we determine this pair of signals to be an astronomical candidate event.

We noted groups of bad pixels with apparent variable brightness, due to defects on the CMOS sensor, so-called ``blinking pixels' \citep{4529048} that caused matches between the two cameras.  For example, a group of nine pixels in Camera B centred at pixel (2251,2524) appeared in a high state with an average duty cycle of $\sim$5\%, which resulted in false matches with Camera A at a rate of $<<$1/hr.  These false matches were simple to identify as they repeated in pixel coordinates and occurred at a low rate.

Many false matches between cameras were seen due to high residuals in the difference imaging associated with the brightest stars to drift through the field.  These false matches were also easy to identify, as they followed the locations of the stars in the east-west direction across sequences of adjacent images, in a predictable manner.

\begin{figure}[!ht]
\begin{center}
\includegraphics[width=0.45\textwidth]{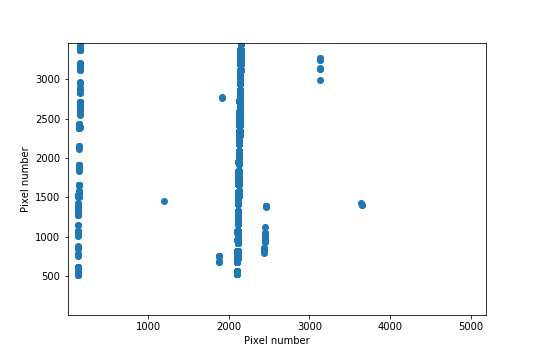}
\caption{Matches between camera A and camera B on 20201011, illustrating the misalignment of the equatorial mount via the inclined trails of matches between differencing residuals corresponding to the passage of bright stars across the sensor over the 3.83 hours of observation at this epoch.  These false positives are easy to identify and disregard from the analysis.}
\label{drift}
\end{center}
\end{figure}

The differencing technique can survive significant misalignment of the equatorial mount.  For example, on 20200722, we set up the system with a purposeful approximate 10$^{\circ}$ misalignment, as a test of the technique.  The processing pipeline performed well under this misalignment, although a higher than normal rate of false positives due to the drift of the brightest stars was noted for these observations (because the drift was not confined only to pixel columns).  Figure \ref{drift} shows the locations of matches between the two cameras for 20201011 (3.83 hours of observations), clearly showing the false positives due to residuals from bright stars, on their inclined tracks across the sensor.  A misalignment would need to result in greater than a column pixel shift between rows to prevent the technique from working completely, which implies a 45$^{\circ}$ rotation of the camera with respect to north.

\begin{figure}[!ht]
\begin{center}
\includegraphics[width=0.22\textwidth]{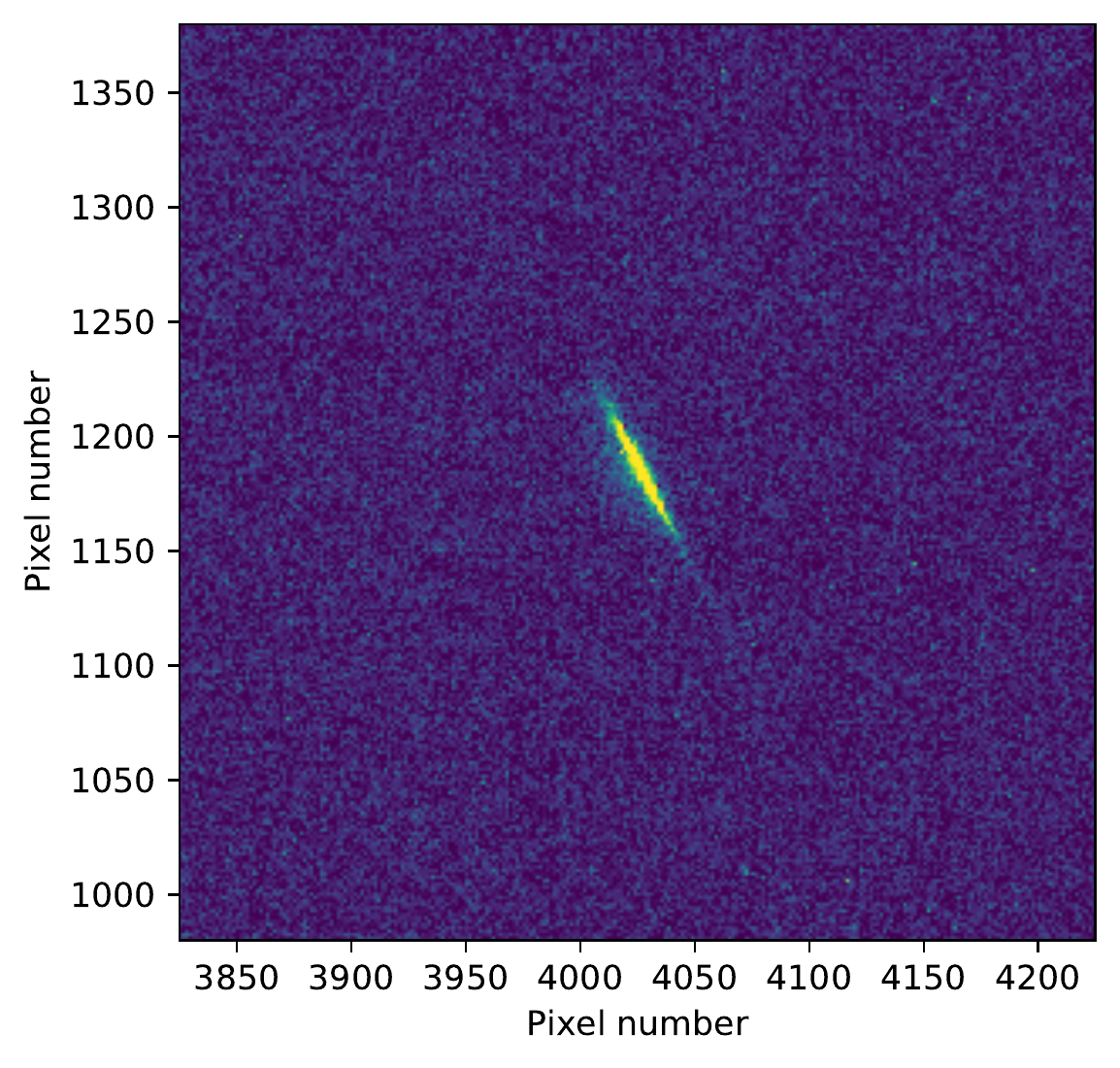}
\includegraphics[width=0.22\textwidth]{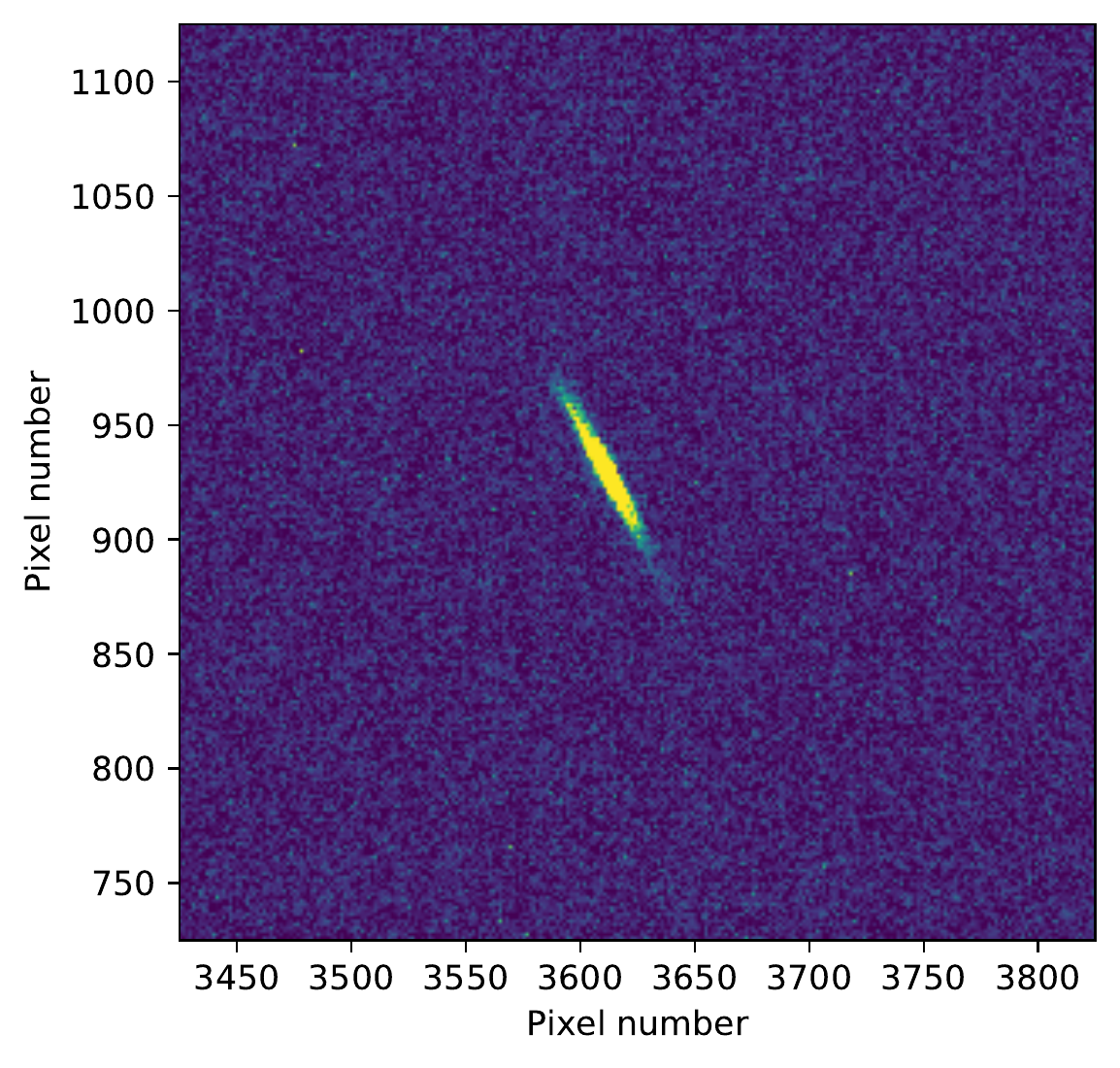}
\caption{Example of a rotating satellite causing a localised flash that triggered the detection pipeline in both cameras (camera A left panel, camera B right panel) at the same celestial coordinates, demonstrating that the system and software work as intended.}
\label{flash}
\end{center}
\end{figure}

Finally, occasional false detections were triggered by satellites.  These were simple to identify, as the satellites produced bright trails through both sets of images that were clearly not astronomical and corresponded to predicted passes of known satellites.  Some such signals produced excellent tests of the system and the data processing.  For example, on 20200722, two images, 25 seconds apart, captured two flashes from a satellite as it rotated.  Both events were identified by our data processing pipeline as candidates, but verified as due to a satellite by a simple inspection of the images. One of these pairs of images is shown in Figure \ref{flash}, as an example.  In this image, the peak brightness of the flash from the satellite corresponds to a magnitude of V$\sim$5.  \citet{2020ApJ...903L..27C} characterise the occurrence of flashes from satellites and measure a peak rate of 1800$\pm^{600}_{280}$ per hour per sky near the equator, at a peak magnitude of 6.8, assuming a flash duration of 0.4 seconds.  This duration is comparable to what we have observed in Figure \ref{flash}.

The subsequent processing steps described in Paper I, determining the astrometry and photometry, need only be performed for any astronomical candidates subsequently confirmed to be high confidence candidates, after a detailed examination of the potential match.

\subsection{Expected false detection rates}
\label{sub:event}
Since cosmic ray hits occur at a rate of approximately one per square centimetre per minute (consistent with the rate seen in Paper I), each camera sensor (3.3 cm$^{2}$ and 5190$\times$3461 pixels) will experience approximately 0.3 cosmic ray hits per exposure.  Thus, the probability of a coincidence between detectors (following filtering of candidates and refinement of astrometry) is $p_{false}=p_{cr}^2(\pi R_{match}^2/N_{pix})$, where $p_{cr}$ is the probability of having a cosmic ray hit in a given image, $R_{match}$ is the matching radius for a match of signals between the two images, and $N_{pix}$ is the total number of pixels in an image.  In words, $p_{false}$ describes the probability of having a cosmic ray hit anywhere on the camera A sensor matched on the camera B sensor within the matching radius.  For our images sizes ($N_{pix}\sim18\times10^{6}$), cosmic ray hit rates (0.3 per camera per exposure), and $R_{match}=10$, $p_{false}\sim10^{-6}$ for any given pair of simultaneous images (assuming every cosmic ray hit generates a signal above the 10$\sigma$ detection threshold).  Over the course of the observations, therefore, the probability of a false astronomical candidate due to cosmic rays is approximately 10$^{-3}$/hour.

Thus, we only expect to require a detailed examination of the images for any (presumably rare) astronomical candidates.  For example, as noted from Paper I, based on Fast Radio Burst (FRB) rates, one FRB might be expected within our field of view every $\sim$70 hours.  The probability of a false astronomical candidate due to cosmic rays in 70 hours is approximately 0.05.

Residual bad pixels not removed by the Bad Pixel Map, for example the intermittent groups of hot pixels noted above, add to this false detection rate, so the false detection rate analysis based on cosmic rays will be a lower limit.  However, the rate of false detections experienced is still greatly reduced.  Over the course of the 24 hours of observations (17,301 pairs of images), approximately only fifty candidate matches required detailed inspection.  None were due to matched cosmic ray hits, as expected from the above analysis; all were due to blinking pixels matched with cosmic ray hits, star trail residuals, or satellite trails/flashes.  Upon individual inspection, all of these candidates could be identified and discarded.

\section{RESULTS AND DISCUSSION}
\label{res}
From our observations covering 89.7 sq. deg. hr, we detect no candidate transient events above our thresholds coincident in celestial coordinates between our two cameras, to an estimated limiting V magnitude of 6.6 and for a cadence of 21 ms.  As noted in Paper I, the corresponding limiting magnitude for a 1 ms transient is approximately 3.3.

Few instruments have entered into the domain of sub-second cadence optical imaging for wide-field blind surveys.  Of note are two experiments.  

The Tomo-e Gozen camera \citep{2020PASJ...72....3R} achieved a 0.5 s cadence over a 1.9 sq. deg. field of view with 1.2'' pixels, and with a 105 cm diameter aperture, giving a limiting magnitude of 15.6 in a 1 s image.  Across 59 hours of exposures over eight nights, no astronomical transients were reported.

We compare our results to those of \citet{2020PASJ...72....3R}, by adopting their method to calculate an upper limit rate per square degree per day at our limiting V magnitude of 6.6.  Using equation 5 from \citet{2020PASJ...72....3R} with the same parameters ($N=0$, $p_{0}=0.05$) but with our aggregate 89.7 sq. deg. hr of observations, we obtain an upper limit for the rate of optical transients at 21 ms duration and V magnitude less than 6.6 of 0.8/sq. deg/day.

\begin{figure}[!ht]
\begin{center}
\includegraphics[width=0.45\textwidth]{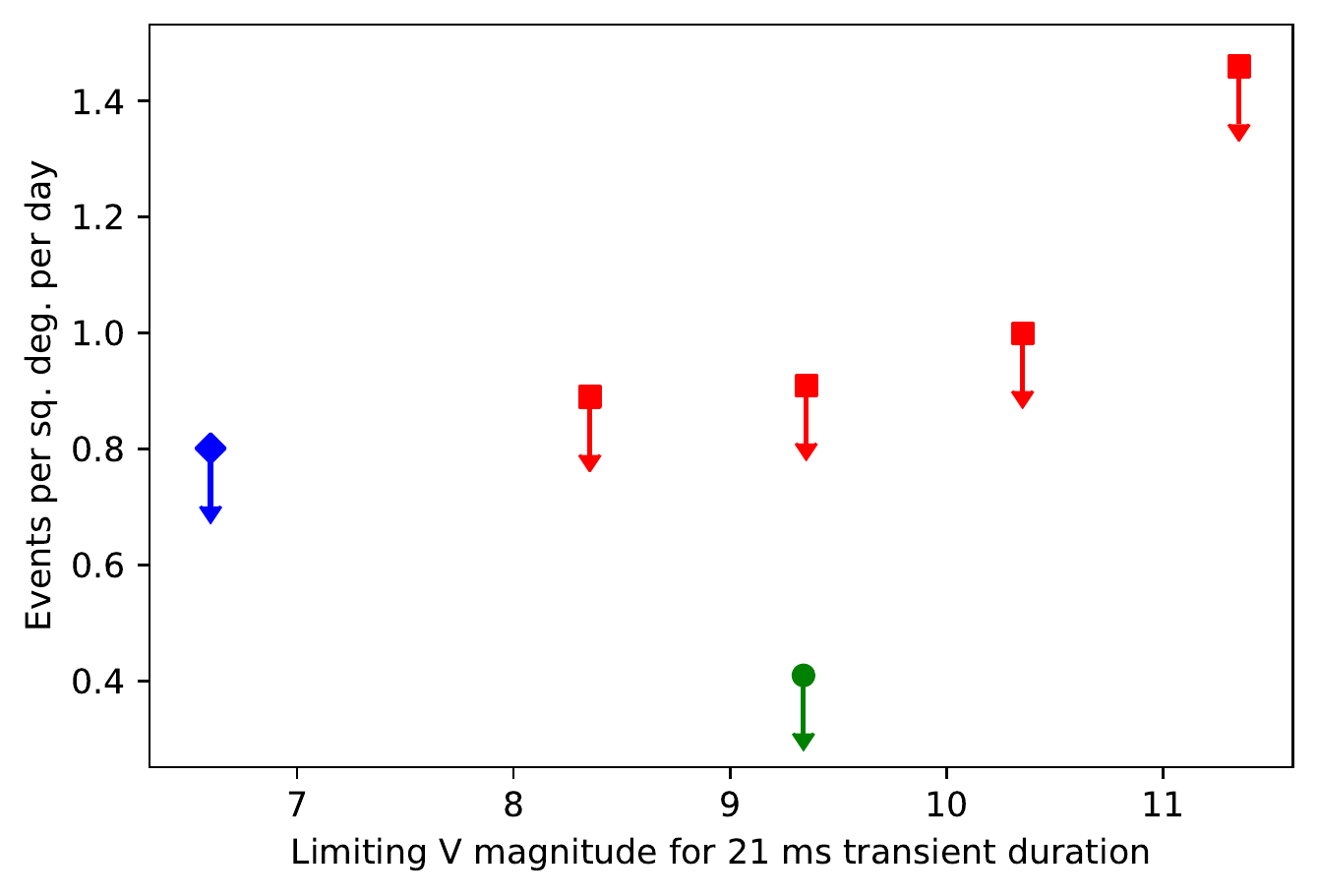}
\caption{Comparison of our results (blue marker, diamond) to the results of \citet{2020PASJ...72....3R}, adjusted to limiting magnitude for 21 ms transient durations (red markers, squares), as described in the text.  The green (circle marker) upper limit is the prediction for the next step system described in Section 5.}
\label{comp}
\end{center}
\end{figure}

In Figure \ref{comp} we compare this result with the results of \citet{2020PASJ...72....3R}, adjusting their results based on 1 second exposure times to the limiting magnitudes expected if the transient duration was 21 ms ($<<$ 1 second).  We can see that the field-of-view improvement of our experiment in comparison to that of \citet{2020PASJ...72....3R} is advantageous.  And even though the raw sensitivity of the Tomo-e Gozen camera significantly exceeds our system, the 1 second exposure limit significantly affects their sensitivity to 21 ms transients, making our results competitive after only 24 hours of accumulated observations.

\citet{2020PASA...37...15T} estimated that Fast Radio Bursts (FRBs) occur at a rate of one event within our sensor solid angle every 70 hours (rate of $\sim0.1$ per sq. deg. per day), so our upper limits do not reach these rates in an absolute sense, irrespective of any expectation regarding the optical magnitudes or durations of events associated with FRBs.

The second experiment of note, the Mini-MegaTORTORA system \citep{2017ASPC..510..526K} provides up to 0.128 s imaging cadence, up to 900 sq. deg. field-of-view, 16'' pixels, an 85 mm diameter aperture, and a limiting V magnitude of 11 in a 0.128 s exposure.  Mini-MegaTORTORA has collected a very large amount of data, being capable of covering the entire sky every observing night, but has not published sub-second results.  The characteristics of Mini-MegaTORTORA should be reasonably favourable for searches of multi-wavelength emission from FRBs.  Chance simultaneous observations of FRBs detected with CHIME \citep{2014SPIE.9145E..22B} may be worth investigating with Mini-MegaTORTORA, although the longitude difference between the telescopes would limit this possibility to circumpolar fields at the Mini-MegaTORTORA latitude.

Both these experiments suffer from the same issues as the single-camera system in Paper I, or as found in \citet{Tingay_2019}, that a short timescale transient that occupies a single image may not be easy to distinguish from cosmic ray events or instrumental effects without additional information.  The Tomo-e Gozen has highly oversampled images, relative to the camera PSF, which means that cosmic ray hits are easier to distinguish from astronomical transients than for critically sampled images.  The notion of oversampling as a method to distinguish between cosmic ray hits and astronomical transients is an example of an instrument imparting a characteristic on an astronomical signal that is not imparted on the cosmic ray hit.  This notion could allow short timescale astronomical transients to be recovered using a single camera.  Other techniques could also be considered, such as running the camera out of focus, although additional effort is then required to recover the sensitivity lost via the image smearing.  Such techniques could be explored in the future with our system.

The two camera system used here provides unambiguous additional information in the form of a coincidence check between the cameras.  Other planned experiments for sub-second optical transient detection may use similar approaches.  The Weizmann Fast Astronomical Survey Telescope (W-FAST; \citet{2017AAS...22915506N}) plans to use two 57 cm f/1.89 telescopes to provide, in the first instance, a 7.5 sq. deg. field of view with 2.3$''$ pixels and 10 ms cadence imaging\footnote{The use of dual telescopes for coincidence testing is not explicitly described by these authors, but is assumed} and later a 23 sq. deg. field.

The Ultra-Fast Astronomy observatory \citep{2019SPIE11341E..1YL} is designed for high time resolution searches for optical emission from repeating FRBs, using dual single-photon resolution fast-response detectors attached to a 70 cm f/6.5 telescope in Kazakhstan.  Previously, \citet{2017MNRAS.472.2800H} and \citet{2020ATel13493....1Z} unsuccessfully searched for optical emission from FRBs at high time resolution, adopting a targeted approach for repeaters rather than a wide-field blind search approach.  

Recent ZTF searches for optical emission from the repeating FRB 180916.J0158$+$65 were mentioned in the introduction, although these observations used 30 s exposures.

So, although the results presented here are based on a modest system and achieve a corresponding modest sensitivity, they represent some of the first meaningful limits on short timescale optical transients with a native cadence $<<$1 second.  The current results demonstrate that a two camera system can efficiently accumulate significant time on sky and filter out false positives from cosmic ray hits on the camera sensors.

\section{CONCLUSIONS AND FUTURE WORK}

\subsection{Conclusions}
We have successfully demonstrated that a simple two camera system using in the drift scanning mode described in Paper I can efficiently place limits on the occurrence of short timescale optical transients.

We find that the two camera system eliminates the false positives caused in a single camera system due to cosmic ray hits on the sensor.  We also uncover, and can reject, other types of false positives at much lower rates, such as the ``blinking pixels'' occasionally seen on the sensors.  Finally, false positives due to differencing residuals at the locations of bright stars are simple to identify and reject.  

Flashes caused by sunlight glints from rotating satellites have provided a test of the system, their coincident detection with both cameras at the same astronomical coordinates a demonstration that the system works as intended.

In the case reported here, with a time resolution of 21 ms and a V magnitude limit of 6.6, we find no astronomical candidate transient events, providing an upper limit of 0.8 events per square degree per day, from a total observation time of approximately 24 hours.

In the context of other similar experiments, this limit is competitive.  The limit is lower than achieved by \citet{2020PASJ...72....3R}, but at a significantly lower limiting magnitude.  However, the \citet{2020PASJ...72....3R} results are for a native time resolution of 1 second, not the sub-second time resolution reported here.  The dual-camera drift scan technique appears promising in order to efficiently probe the question of the rate of occurrence of short timescale optical transients that may be produced by a variety of physical mechanisms and associated with events at other wavelengths, such as FRBs.

\subsection{Future work}
The work reported here completes the successful extension of the pre-prototype system described in Paper I.

The next step for the development of the technique is to move to larger apertures and wider fields of view.  As flagged in Paper I, this will be achieved by continuing the off-the-shelf approach.  We plan to utilise two Celestron 279 mm aperture, f/2.2 Rowe-Ackermann Astrographs (620 mm focal length).  An efficient choice for the cameras would be the Canon EOS R, with a ``full frame'' CMOS sensor of 36 mm $\times$ 24 mm providing a 7.3 sq. deg. field-of-view with the astrograph.  Using such a system from the same location, for the same time resolution and observing time, would result in the predicted improvement shown in Figure \ref{comp}.  Such a system would demand a dark sky location, in which case the improvement is likely to be better than indicated in Figure \ref{comp} (in terms of limiting magnitude).

As a larger and heavier system, some significant engineering will be required to realise it.  This new system will be realised during 2021.



\begin{acknowledgements}
We thank the anonymous referee for their positive review and encouragement, and for comments that improved the paper.  This research has made use of the NASA/IPAC Extragalactic Database (NED), which is funded by the National Aeronautics and Space Administration and operated by the California Institute of Technology.
\end{acknowledgements}

\bibliographystyle{pasa-mnras}
\bibliography{custom}

\begin{thebibliography}{}
\makeatletter
\relax
\def\mn@urlcharsother{\let\do\@makeother \do\$\do\&\do\#\do\^\do\_\do\%\do\~}
\definecolor{darkblue}{rgb}{0,0,0.597656}
\def\mndoi{\begingroup\mn@urlcharsother \@ifnextchar [ {\mndoi@} {\mndoi@[]}}
\def\mndoi@[#1]#2{\def\@tempa{#1}\ifx\@tempa\@empty \href
  {http://dx.doi.org/#2} {\textcolor{darkblue}{doi:#2}}\else \href
  {http://dx.doi.org/#2} {\textcolor{darkblue}{#1}}\fi \endgroup}
\def\mn@eprint#1#2{\mn@eprint@#1:#2::\@nil}
\def\mn@eprint@arXiv#1{\href {http://arxiv.org/abs/#1} {{\tt arXiv:#1}}}
\def\mn@eprint@dblp#1{\href {http://dblp.uni-trier.de/rec/bibtex/#1.xml}
  {dblp:#1}}
\def\mn@eprint@#1:#2:#3:#4\@nil{\def\@tempa {#1}\def\@tempb {#2}\def\@tempc
  {#3}\ifx \@tempc \@empty \let \@tempc \@tempb \let \@tempb \@tempa \fi \ifx
  \@tempb \@empty \def\@tempb {arXiv}\fi \@ifundefined
  {mn@eprint@\@tempb}{\@tempb:\@tempc}{\expandafter \expandafter \csname
  mn@eprint@\@tempb\endcsname \expandafter{\@tempc}}}

\bibitem[\protect\citeauthoryear{{Ackerson}, {Musante}, {Gambino},
  {Ellis-Monaghan}, {Maynard}, {Rassel}, {Ogg}  \& {Jaffe}}{{Ackerson}
  et~al.}{2008}]{4529048}
{Ackerson} K.,  {Musante} C.,  {Gambino} J.,  {Ellis-Monaghan} J.,  {Maynard}
  D.,  {Rassel} R.~J.,  {Ogg} K.,   {Jaffe} M.,  2008, in 2008 IEEE/SEMI
  Advanced Semiconductor Manufacturing Conference. pp 255--258

\bibitem[\protect\citeauthoryear{{Andreoni} et~al.,}{{Andreoni}
  et~al.}{2020}]{2020arXiv200506273A}
{Andreoni} I.,  et~al., 2020, arXiv e-prints, \href
  {https://ui.adsabs.harvard.edu/abs/2020arXiv200506273A} {p. arXiv:2005.06273}

\bibitem[\protect\citeauthoryear{{Bandura} et~al.,}{{Bandura}
  et~al.}{2014}]{2014SPIE.9145E..22B}
{Bandura} K.,  et~al., 2014, {Canadian Hydrogen Intensity Mapping Experiment
  (CHIME) pathfinder}.
p. 914522, \mndoi{10.1117/12.2054950}

\bibitem[\protect\citeauthoryear{{Burke-Spolaor}}{{Burke-Spolaor}}{2018}]{2018NatAs...2..845B}
{Burke-Spolaor} S.,  2018, \mndoi [Nature Astronomy]
  {10.1038/s41550-018-0630-x}, \href
  {https://ui.adsabs.harvard.edu/abs/2018NatAs...2..845B} {2, 845}

\bibitem[\protect\citeauthoryear{{Corbett} et~al.,}{{Corbett}
  et~al.}{2020}]{2020ApJ...903L..27C}
{Corbett} H.,  et~al., 2020, \mndoi [\apjl] {10.3847/2041-8213/abbee5}, \href
  {https://ui.adsabs.harvard.edu/abs/2020ApJ...903L..27C} {903, L27}

\bibitem[\protect\citeauthoryear{{Hardy} et~al.,}{{Hardy}
  et~al.}{2017}]{2017MNRAS.472.2800H}
{Hardy} L.~K.,  et~al., 2017, \mndoi [\mnras] {10.1093/mnras/stx2153}, \href
  {https://ui.adsabs.harvard.edu/abs/2017MNRAS.472.2800H} {472, 2800}

\bibitem[\protect\citeauthoryear{{Karpov} et~al.,}{{Karpov}
  et~al.}{2017}]{2017ASPC..510..526K}
{Karpov} S.,  et~al., 2017, {Mini-MegaTORTORA Wide-Field Monitoring System with
  Subsecond Temporal Resolution: Observation of Transient Events}.
p.~526

\bibitem[\protect\citeauthoryear{{Li}, {Smoot}, {Grossan}, {Lau},
  {Bekbalanova}, {Shafiee}  \& {Stezelberger}}{{Li}
  et~al.}{2019}]{2019SPIE11341E..1YL}
{Li} S.,  {Smoot} G.~F.,  {Grossan} B.,  {Lau} A. W.~K.,  {Bekbalanova} M.,
  {Shafiee} M.,   {Stezelberger} T.,  2019, in \procspie. p. 113411Y
  (\mn@eprint {arXiv} {1908.10549}), \mndoi{10.1117/12.2548169}

\bibitem[\protect\citeauthoryear{{Nir}, {Ofek}, {Ben-Ami}, {Manulis},
  {Gal-Yam}, {Diner}  \& {Rappaport}}{{Nir} et~al.}{2017}]{2017AAS...22915506N}
{Nir} G.,  {Ofek} E.~O.,  {Ben-Ami} S.,  {Manulis} I.,  {Gal-Yam} A.,  {Diner}
  O.,   {Rappaport} M.,  2017, in American Astronomical Society Meeting
  Abstracts \#229. p. 155.06

\bibitem[\protect\citeauthoryear{{Richmond} et~al.,}{{Richmond}
  et~al.}{2020}]{2020PASJ...72....3R}
{Richmond} M.~W.,  et~al., 2020, \mndoi [\pasj] {10.1093/pasj/psz120}, \href
  {https://ui.adsabs.harvard.edu/abs/2020PASJ...72....3R} {72, 3}

\bibitem[\protect\citeauthoryear{{Tingay}}{{Tingay}}{2020}]{2020PASA...37...15T}
{Tingay} S.,  2020, \mndoi [\pasa] {10.1017/pasa.2020.7}, \href
  {https://ui.adsabs.harvard.edu/abs/2020PASA...37...15T} {37, e015}

\bibitem[\protect\citeauthoryear{Tingay \& Yang}{Tingay \&
  Yang}{2019}]{Tingay_2019}
Tingay S.~J.,  Yang Y.-P.,  2019, \mndoi [The Astrophysical Journal]
  {10.3847/1538-4357/ab2c6e}, 881, 30

\bibitem[\protect\citeauthoryear{{Zampieri} et~al.,}{{Zampieri}
  et~al.}{2020}]{2020ATel13493....1Z}
{Zampieri} L.,  et~al., 2020, The Astronomer's Telegram, \href
  {https://ui.adsabs.harvard.edu/abs/2020ATel13493....1Z} {13493, 1}

\makeatother
\end{thebibliography}

\end{document}